\documentclass[aps,prl,twocolumn,nofootinbib,
superscriptaddress,showpacs,preprintnumbers,altaffilletter,10pt]{revtex4-1}

\usepackage{amsmath,amsfonts,amssymb}
\usepackage{graphicx,bm,color,subfigure,hyperref,xspace}
\usepackage[load=addn]{siunitx}
\usepackage{changes}
\usepackage{lineno}
\usepackage[percent]{overpic}
\usepackage{csquotes}

\begin{document}

%%%%%%%%%%%%%%%%%%%%%%%%%%%%%%%%%%%%%%%%%%%%%%%%%%%%%%%%%%
% Title
%%%%%%%%%%%%%%%%%%%%%%%%%%%%%%%%%%%%%%%%%%%%%%%%%%%%%%%%%%
\title{Triangle Singularity as the Origin of the $a_1(1420)$}

%%%%%%%%%%%%%%%%%%%%%%%%%%%%%%%%%%%%%%%%%%%%%%%%%%%%%%%%%%
% Author list
%%%%%%%%%%%%%%%%%%%%%%%%%%%%%%%%%%%%%%%%%%%%%%%%%%%%%%%%%%
\affiliation{Department of Physics, University of Aveiro, I3N, 3810-193 Aveiro, Portugal}
\affiliation{Institut f\"ur Experimentalphysik, Universit\"at Bochum, 44780 Bochum, Germany}
\affiliation{Helmholtz-Institut f\"ur  Strahlen- und Kernphysik, Universit\"at Bonn, 53115 Bonn, Germany}
\affiliation{Physikalisches Institut, Universit\"at Bonn, 53115 Bonn, Germany}
\affiliation{Institute of Scientific Instruments of the CAS, 61264 Brno, Czech Republic}
\affiliation{Matrivani Institute of Experimental Research \& Education, Calcutta-700 030, India}
\affiliation{Joint Institute for Nuclear Research, 141980 Dubna, Moscow region, Russia}
\affiliation{Physikalisches Institut, Universit\"at Freiburg, 79104 Freiburg, Germany}
\affiliation{CERN, 1211 Geneva 23, Switzerland}
\affiliation{Technical University in Liberec, 46117 Liberec, Czech Republic}
\affiliation{LIP, 1649-003 Lisbon, Portugal}
\affiliation{Institut f\"ur Kernphysik, Universit\"at Mainz, 55099 Mainz, Germany}
\affiliation{University of Miyazaki, Miyazaki 889-2192, Japan}
\affiliation{Lebedev Physical Institute, 119991 Moscow, Russia}
\affiliation{Physik Department, Technische Universit\"at M\"unchen, 85748 Garching, Germany}
\affiliation{Nagoya University, 464 Nagoya, Japan}
\affiliation{Faculty of Mathematics and Physics, Charles University, 18000 Prague, Czech Republic}
\affiliation{Czech Technical University in Prague, 16636 Prague, Czech Republic}
\affiliation{State Scientific Center Institute for High Energy Physics of National Research Center "Kurchatov Institute", 142281 Protvino, Russia}
\affiliation{IRFU, CEA, Universit\'e Paris-Saclay, 91191 Gif-sur-Yvette, France}
\affiliation{Academia Sinica, Institute of Physics, Taipei 11529, Taiwan}
\affiliation{School of Physics and Astronomy, Tel Aviv University, 69978 Tel Aviv, Israel}
\affiliation{Department of Physics, University of Trieste, 34127 Trieste, Italy}
\affiliation{Trieste Section of INFN, 34127 Trieste, Italy}
\affiliation{Department of Physics, University of Torino, 10125 Turin, Italy}
\affiliation{Torino Section of INFN, 10125 Turin, Italy}
\affiliation{Tomsk Polytechnic University, 634050 Tomsk, Russia}
\affiliation{Department of Physics, University of Illinois at Urbana-Champaign, Urbana, IL 61801-3080, USA}
\affiliation{National Centre for Nuclear Research, 02-093 Warsaw, Poland}
\affiliation{Faculty of Physics, University of Warsaw, 02-093 Warsaw, Poland}
\affiliation{Institute of Radioelectronics, Warsaw University of Technology, 00-665 Warsaw, Poland}
\affiliation{Yamagata University, Yamagata 992-8510, Japan}
\author{G.~D.~Alexeev}
\affiliation{Joint Institute for Nuclear Research, 141980 Dubna, Moscow region, Russia}
\author{M.~G.~Alexeev}
\affiliation{Department of Physics, University of Torino, 10125 Turin, Italy}
\affiliation{Torino Section of INFN, 10125 Turin, Italy}
\author{A.~Amoroso}
\affiliation{Department of Physics, University of Torino, 10125 Turin, Italy}
\affiliation{Torino Section of INFN, 10125 Turin, Italy}
\author{V.~Andrieux}
\affiliation{CERN, 1211 Geneva 23, Switzerland}
\affiliation{Department of Physics, University of Illinois at Urbana-Champaign, Urbana, IL 61801-3080, USA}
\author{V.~Anosov}
\affiliation{Joint Institute for Nuclear Research, 141980 Dubna, Moscow region, Russia}
\author{A.~Antoshkin}
\affiliation{Joint Institute for Nuclear Research, 141980 Dubna, Moscow region, Russia}
\author{K.~Augsten}
\affiliation{Joint Institute for Nuclear Research, 141980 Dubna, Moscow region, Russia}
\affiliation{Czech Technical University in Prague, 16636 Prague, Czech Republic}
\author{W.~Augustyniak}
\affiliation{National Centre for Nuclear Research, 02-093 Warsaw, Poland}
\author{C.~D.~R.~Azevedo}
\affiliation{Department of Physics, University of Aveiro, I3N, 3810-193 Aveiro, Portugal}
\author{B.~Bade{\l}ek}
\affiliation{Faculty of Physics, University of Warsaw, 02-093 Warsaw, Poland}
% ...

%%%%%%%%%%%%%%%%%%%%%%%%%%%%%%%%%%%%%%%%%%%%%%%%%%%%%%%%%%
\newcommand{\diff}{\ensuremath{\mathrm{d}}}
\newcommand{\ie}{\textit{i.e.}\xspace}
\newcommand{\eg}{\textit{e.g.}\xspace}
\newcommand{\cf}{\textit{cf.}\xspace}
\newcommand{\onePPfnotP}{\ensuremath{1^{++}\,0^+\, f_0\pi\, P}}
\newcommand{\fnotP}{\ensuremath{f_0\pi\, P}}
\newcommand{\rhoS}{\ensuremath{\rho\pi\, S}}
\newcommand{\rhoD}{\ensuremath{\rho\pi\, D}}
\newcommand{\Mat}{\tilde{\mathcal{M}}}
\newcommand{\R}{\ensuremath{\mathcal{R}}}
\newcommand{\Rred}{\ensuremath{\mathcal{R}^2_\text{red}}}
\newcommand{\Po}{\ensuremath{I\!\!P}}

%%%%%%%%%%%%%%%%%%%%%%%%%%%%%%%%%%%%%%%%%%%%%%%%%%%%%%%%%%

%%%%%%%%%%%%%%%%%%%%%%%%%%%%%%%%%%%%%%%%%%%%%%%%%%%%%%%%%%
% Abstract
%%%%%%%%%%%%%%%%%%%%%%%%%%%%%%%%%%%%%%%%%%%%%%%%%%%%%%%%%%

\begin{abstract}
The COMPASS experiment recently discovered a new isovector resonance-like signal with axial-vector quantum numbers, the $a_1(1420)$, decaying to $f_0(980)\pi$. With a mass too close to and a width smaller than the axial-vector ground state $a_1(1260)$, it was immediately interpreted as a new light exotic meson, similar to the $X$, $Y$, $Z$ states in the hidden-charm sector.
We show that a resonance-like signal fully matching the experimental data is produced by the decay of the $a_1(1260)$ resonance into
$K^\ast(\to K\pi)\bar{K}$ and subsequent rescattering through a triangle singularity into the coupled $f_0(980)\pi$ channel.
The amplitude for this process is calculated using a new approach based on dispersion relations.
The triangle-singularity 
model is fitted to the partial-wave data of the COMPASS experiment.
Despite having less parameters, this fit shows a slightly better quality than the one using a
resonance hypothesis and thus eliminates the need for an additional resonance in order to describe the data.
We thereby demonstrate for the first time in the light-meson sector that a resonance-like structure in the experimental data can be described by rescattering through a triangle singularity, providing evidence for a genuine three-body effect.
\end{abstract}

\maketitle

%%%%%%%%%%%%%%%%%%%%%%%%%%%%%%%%%%%%%%%%%%%%%%%%%%%%%%%%%%
% Introduction
%%%%%%%%%%%%%%%%%%%%%%%%%%%%%%%%%%%%%%%%%%%%%%%%%%%%%%%%%%

Quantum chromodynamics is generally accepted as the fundamental quantum-field theory of the strong interaction.
How exactly the spectrum of bound states (hadrons) emerges from the underlying interaction between quarks and gluons is, however, not yet understood. The main difficulty is the rise of the strong coupling at the low-energy scale relevant for hadrons, which makes the theory unsolvable with perturbative methods.
Although the constituent-quark model~\cite{Tanabashi:2018oca,Ebert:2009ub,Eichten:1978tg} describes many of the observed mesons, it seems that the spectrum is notably richer:
there is growing experimental evidence for bound states beyond the constituent-quark model. Such states are commonly called exotic~\cite{Ketzer:2012vn,Meyer:2015eta,Esposito:2016noz,Guo:2017jvc,Olsen:2017bmm,Rodas:2018owy}.
In addition to mapping out the full spectrum predicted by models and, more recently, by lattice gauge theory \cite{Dudek:2013yja}, the search for such exotic states drives the current interest in hadron spectroscopy.

The study of single-diffractive reactions with a high-energy meson beam, as
performed by the COMPASS experiment at the CERN SPS~\cite{Abbon:2007pq,Abbon:2014aex},
is a natural way to investigate
meson excitations (for a recent review, see~\cite{Ketzer:2019wmd}).
In such
reactions, at high energies commonly described by the exchange of a Pomeron $\Po$, 
the incoming beam particle is
excited by the strong interaction with a proton target.
Regge theory then allows us to factorize off the target vertex, such that we only consider the beam vertex.
Although the produced excited system immediately decays, the reaction products unveil the properties of the excitation.
An unprecedented amount of data comprising
almost $50\,$ million events  
for the reaction $\pi^-+p\to \pi^-\pi^-\pi^+ + p$
were used by COMPASS to perform a detailed analysis 
of $\pi_J$ and $a_J$ mesons with isospin $I=1$, negative $G$-parity, and positive $C$-parity implied by $G=C (-1)^I$.
The partial-wave analysis (PWA) technique in connection with the isobar model was used to separate excitations with different quantum numbers, see~\cite{Ketzer:2019wmd,Adolph:2015tqa} for details.
Individual waves are labeled
$J^{PC}\,M^\epsilon\,\xi\pi\,L$, where $J$ 
is the total angular momentum of the 3-pion system,
$P$ the spatial and $C$ the charge-conjugation parity.
The quantum number $M$ labels the projection of the spin $J$ onto the direction of the beam in the rest frame of $\pi^-\pi^-\pi^+$, and $\epsilon$ indicates the reflection symmetry with respect to the production plane.
At the high center-of-momentum energies of the experiment, the reflectivity quantum number $\epsilon$ corresponds to the naturality of the exchanged particle and is hence always positive for Pomeron exchange. 
The orbital angular momentum between the neutral system of two pions (isobar) and the remaining pion
is denoted by $L$.
The symbol $\xi$ labels the assumed isobar, \ie the interaction amplitude in the neutral $\pi\pi$-subchannel.

\begin{figure*}[tbp]
    \centering
    \begin{overpic}[width=.44\textwidth]{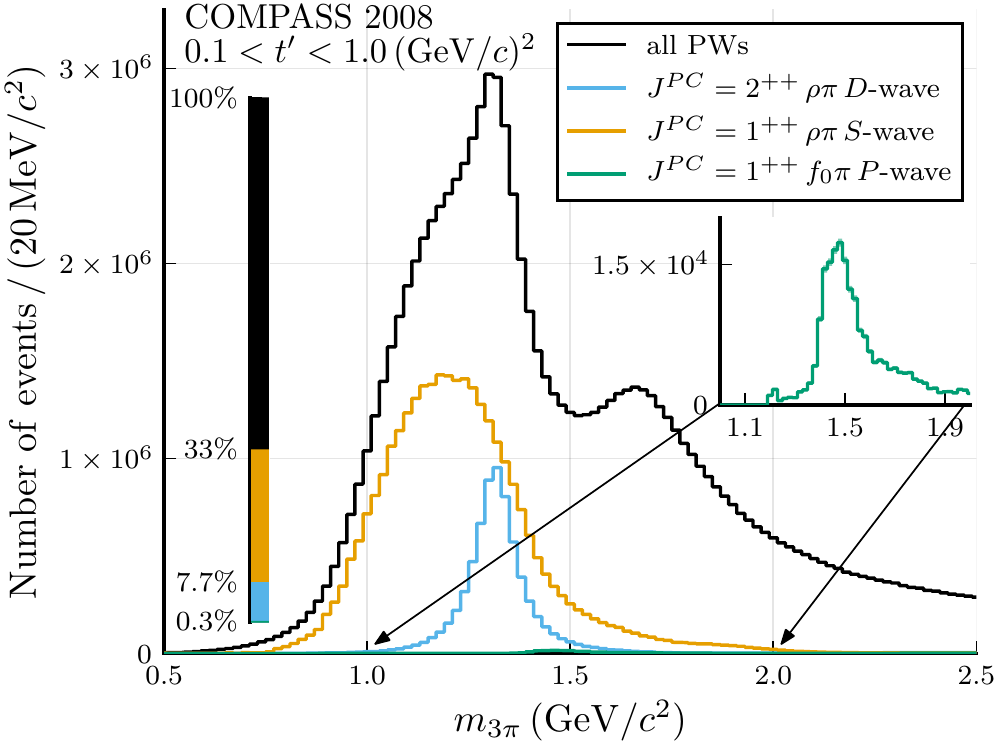} \put (30,62) {(a)}\end{overpic}
    \begin{overpic}[width=.52\textwidth]{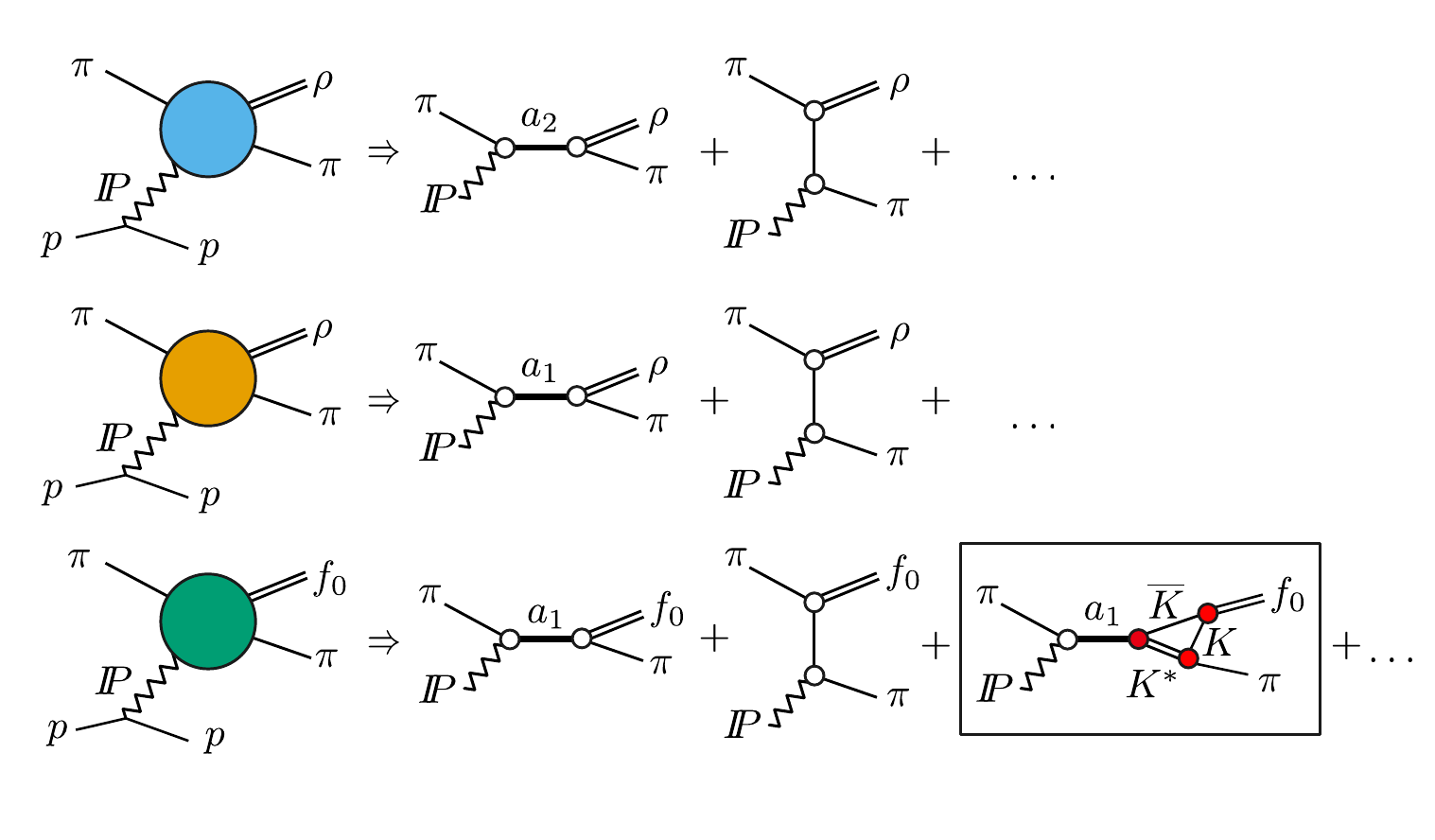} \put (80,55) {(b)}\end{overpic}
    \caption{(a)
    Intensities of selected waves from the
    PWA of the reaction $\pi^-+p\to \pi^-\pi^-\pi^+ +p$~\cite{Adolph:2015tqa}.
    The inset shows an enlarged view of the $1^{++}0^+\,f_0 \pi\,P$-wave.
    The colored bar on the left indicates the contributions of the different waves to the total intensity.
    (b) Diagrams showing possible contributions to the $\rho(770)\pi$ and $f_0(980)\pi$ production amplitudes.
    The Pomeron is labeled $\Po$, $a_1$ refers to the axial-vector ground state $a_1(1260)$, and $a_2$ to the tensor ground state $a_2(1320)$. 
    The framed diagram shows the dominant contribution to the $a_1(1420)$ signal via the triangle diagram.
    \label{fig:pwa}
    }
\end{figure*}

A PWA including $88$ waves in total was performed separately in 100 bins of the $3\pi$ invariant mass $m_{3\pi}$ and 11 bins of the reduced 4-momentum transfer squared $t'$ (see Eq.~(6) in~\cite{Adolph:2015tqa}).
The results are summarized in Fig.~\ref{fig:pwa}(a),
where we show the intensities of selected waves as a function of $m_{3\pi}$, summed over all $t'$ bins.
Among many important observations, an exotic resonance-like signal with quantum numbers $J^{PC}=1^{++}$ was found in the \onePPfnotP-wave as a clear peak at $1.4\,\mathrm{GeV}/c^2$~\cite{Adolph:2015pws}
(see inset of Fig.~\ref{fig:pwa}(a)).
The resonance-like behavior was 
corroborated by the
observed phase motion, \ie a mass-dependent relative phase with respect to several other reference waves.
Extensive studies, also using the ``freed-isobar" method~\cite{Krinner:2017dba}, undoubtedly confirmed the signal and proved that it was not an artifact of any particular isobar parametrization~\cite{Adolph:2015tqa}.
Following the PDG convention, the signal was called $a_1(1420)$ according to its quantum numbers $I^G (J^{PC})=1^- (1^{++})$. 
It was immediately realized that it 
could not be an ordinary quark-model meson resonance:
(i) with about $150\,\mathrm{MeV}/c^2$, its width is much smaller than that
of the axial-vector ground state $a_1(1260)$
of about $500\,\mathrm{MeV}/c^2$; 
(ii) the signal is separated from the ground state by only about $150\,\mathrm{MeV}/c^2$,
whereas
the energy difference between different radial excitation levels
is typically $400\,\mathrm{MeV}/c^2$ as estimated 
based on the slope of the radial excitation trajectory~\cite{Chen:2015iqa,Anisovich:2000kxa};
(iii) so far, the $a_1(1420)$ was seen only in the $f_0(980)\pi$ final state.

Various interpretations followed the $a_1(1420)$ observation ~\cite{Ketzer:2015tqa,Aceti:2016yeb,Chen:2015fwa,Gutsche:2017twh,Basdevant:2015wma},
requiring or not a new resonance.
Resonances are consistently introduced in general scattering theory~\cite{Gribov:2009zz},
where the reaction amplitude is an analytic function of the total energy squared $s$
that is regarded as a complex number; they
are found as poles on the unphysical sheet of the complex $s$-plane attached to the real axis from below.
In explanations involving either diquark-antidiquark molecules or tightly bound tetraquarks, the observed signal, \ie peak
and phase motion, is caused by a pole-type singularity located on the closest sheet.
Alternatively, a so-called triangle-singularity (TS) mechanism~\cite{Gribov:2009zz,Eden:1971jm,Guo:2019twa}
was proposed as the mechanism behind the $a_1(1420)$ signal~\cite{Ketzer:2015tqa}. 
Here, a logarithmic branch point caused by
a coupled-channel effect, particularly by the $K^\ast\bar{K}$-$f_0\pi$ interaction, is
located near the physical region on the closest unphysical sheet. 
The other proposed model~\cite{Basdevant:2015wma,Basdevant:1977ya} that
does not require a new resonance pole,
combines resonant and nonresonant production mechanisms
resulting in a peak in the \onePPfnotP-wave.
However, the generated phase motion is
at the position of the $a_1(1260)$ resonance, which is
inconsistent with observation.

In this Letter, we interpret the COMPASS data in terms of
the triangle-singularity model based on a new method for the calculation of the amplitude.
The calculation implements the proposal of~\cite{Mikhasenko:2019vhk} exploiting the unitarity and analyticity properties of the amplitude.
The new model goes beyond~\cite{Ketzer:2015tqa} by incorporating spin in a more systematic way and
allowing us to address higher-order rescattering effects.
To our knowledge, this is the first time that the TS model, mimicking a resonance signal, is fitted successfully to experimental data in the light-meson sector describing both intensity and phase motion simultaneously.
Comparable studies in the heavy-quark sector, see e.g.~\cite{Szczepaniak:2015eza, Guo:2017jvc, Nakamura:2019btl}, were performed on a much smaller statistical basis.

%%%%%%%%%%%%%%%%%%%%%%%%%%%%%%%%%%%%%%%%%%%%%%%%%%%%%%%%%%
% Rescattering effects in diffractive production
%%%%%%%%%%%%%%%%%%%%%%%%%%%%%%%%%%%%%%%%%%%%%%%%%%%%%%%%%%
The dynamics of a hadronic three-body system is commonly understood in terms of quasi-two-body interactions with subchannel resonances $\xi$ decaying further into pairs of final-state particles.
Often, however, the same final state can be obtained through several decay chains when the two-particle interaction is non-negligible for different particle pairs~\cite{Herndon:1973yn,Aitchison:1965zz}.
Different decay chains are coherent, hence they interfere. The unitarity of the scattering matrix enforces a consistency relation between the different chains~\cite{Khuri:1960zz,Pasquier:1968zz,Pasquier:1969dt}.
This relation makes the line shapes of the resonances in a particle pair in a system of three particles dependent on the dynamics in the other pairs~\cite{Aitchison:1979fj,Niecknig:2015ija,Niecknig:2012sj,Danilkin:2014cra}.
An equivalent way of describing this interrelation between pair-wise interactions is to state
that the cross-channel two-body resonances in the $\pi\pi$, $K\pi$ and $\bar{K}K$ systems 
rescatter to one another, 
thereby modifying the original undistorted line shapes.
In addition, the probabilities for a three-body resonance decaying to one or another channel may be redistributed due to 
final-state interaction~\cite{Schmid:1967ojm,Szczepaniak:2015hya}. The latter effect is strongly enhanced for certain kinematic conditions~\cite{Landau:1959fi,Gribov:2009zz} and produces the observed resonance-like signal in the case considered here. 

We find that the presence of the $K^\ast(892)$ resonance (hereafter referred to as $K^\ast$)
in the $K\bar{K}\pi$ channel drastically affects the $f_0(980)\pi$ channel, since the rescattering between $K^\ast\to K\pi\,P$-wave and $K\bar{K}\to f_0(980)\to \pi\pi\,S$-wave
occurs with all intermediate particles being almost on their mass shell for
$m_{3\pi}\approx 1.4\,$GeV$/c^2$, \ie~slightly above the $K^\ast\bar{K}$ 
threshold~\cite{Ketzer:2015tqa}.
This effect does not disturb the narrow line shape of the $f_0(980)$, but it leads to a significant redistribution of the $a_1(1260)$ decay probabilities.
The originally negligible $f_0(980)\pi\,P$-wave decay channel is populated by the rescattering from the $K^\ast\bar{K}$ decay locally around $1.4\,$GeV$/c^2$.

%%%%%%%%%%%%%%%%%%%%%%%%%%%%%%%%%%%%%%%%%%%%%%%%%%%%%%%%%%
% Calculation of TS
%%%%%%%%%%%%%%%%%%%%%%%%%%%%%%%%%%%%%%%%%%%%%%%%%%%%%%%%%%
Our calculation of the TS amplitude is reminiscent of the Khuri-Treiman (KT) equation first developed in 1960~\cite{Khuri:1960zz,Pasquier:1968zz,Pasquier:1969dt}:
the dispersion relation and two-body unitarity are used to connect the isobar amplitude with the partial-wave projection of the cross channels.
By $F_W^{\{1\}}$ we denote the production amplitude  of a three-particle system $(123)$ with a given set of quantum numbers $W$, the invariant mass squared $s\equiv m_{3\pi}^2$, and the isobar formed by particles $2$ and $3$ (hereafter labeled $\{1\}$ using indices in curly brackets) with invariant mass squared $\sigma_{\{1\}} \equiv (p_2+p_3)^2$.
We write the dispersion relation for the kinematic-singularity-free amplitude
$F_W^{\{1\}}(s,\sigma_{\{1\}}) / K^{\{1\}}_W(s,\sigma_{\{1\}})$
with $K^{\{1\}}_W$ being the break-up momentum for the $f_0 \pi$ system required for the $P$-wave~(see Supplemental Material):
\begin{align} \label{eq:rescattering}
  &F_W^{\{1\}}(s,\sigma_{\{1\}}) = K^{\{1\}}_W(s,\sigma_{\{1\}}) \bigg( C_W^{\{1\}}(s) \\ \nonumber
    & \quad 
    + \frac{1}{2\pi} \sum_\omega \int_{\sigma_{\text{th},W}}^\infty \frac{\rho_\omega(\sigma) \hat F_{W,\omega}^{\{1\}}(s,\sigma)}{K^{\{1\}}_W(s,\sigma)(\sigma - \sigma_{\{1\}}-i\varepsilon)} \text{d}\sigma \bigg). 
\end{align}
\noindent Here, the indices $W$ and $\omega$ refer to the full set of quantum numbers labeling a given wave.
For the \fnotP-wave considered in this paper,  
the term $C_W^{\{1\}}(s)$ parametrizes the three-body production dynamics and the decay into the given final state $W$.
It includes the direct production of the $a_1(1260)$ resonance and
a term for the nonresonant production that is further described below.
The sum runs over all possible cross channels with quantum numbers $\omega$. In the dispersion integral, $\rho_\omega(\sigma)$ is the 2-body phase-space factor, and $\hat{F}_{W,\omega}^{\{1\}}(s,\sigma)$ is the projection of the cross channel ${\{3\}}$, i.e.~the
isobar formed by particles $1$ and $2$, with quantum numbers $\omega$ onto channel $\{1\}$ with quantum numbers $W$.
We do not expect isobars in channel $\{2\}$, formed by, \eg, $K^-\pi^-$, $K^0\pi^-$, or $\pi^-\pi^-$.
We note an important difference to the original KT equation.
The latter constrains the subchannel dynamics in the three-body system.
The total invariant mass of the system is treated as a fixed parameter in the model.
In 1965, Aitchison suggested that this parametric dependence is actually physical and represents
the three-body interaction~\cite{Aitchison:1965zz}.
In~\cite{Mikhasenko:2019vhk}, the authors demonstrated that the KT kernel can be used to separate
the genuine three-body dynamics from the final-state interaction.
Correspondingly, in Eq.~\eqref{eq:rescattering} the direct decay of $a_1(1260)$ enters in $C_W^{\{1\}}(s)$, while the $s$-dependent dispersion integral adds the rescattering corrections.
Assuming that modifications of the line shapes of the cross-channel resonances due to rescattering are negligible, we find that the $K^\ast\bar{K}$ channel produces a narrow peak and a strong phase motion at the mass of the $a_1(1420)$ due to the TS being very close to the physical region, while all other possible rescattering corrections, which we investigated, manifest themselves in a broad bump and a slow phase motion similar to the direct decay and the nonresonant background (see Supplemental Material).

%%%%%%%%%%%%%%%%%%%%%%%%%%%%%%%%%%%%%%%%%%%%%%%%%%%%%%%%%%
% Fit model => moved after the formalism
%%%%%%%%%%%%%%%%%%%%%%%%%%%%%%%%%%%%%%%%%%%%%%%%%%%%%%%%%%

For a fit of the TS model to the COMPASS spin-density matrix elements~\cite{paper3_hepdata},
referred to below as the data points, we choose the three waves depicted in Fig.~\ref{fig:pwa}(b), which constitute the dominant contributions to the $\rho\pi$ and $f_0\pi$ production amplitudes: (i) the $1^{++}0^+\,\rho\pi\, S$-wave describes the source of the rescattering process, since its largest contribution comes from the $a_1(1260)$. This wave also contains a significant contribution from nonresonant ``Deck''-like processes~\cite{Ascoli:1974sp}; 
(ii) the $1^{++}0^+\,f_0\pi\,P$-wave contains the $a_1(1420)$ signal;
(iii) the $2^{++}1^+\,\rho\pi\, D$-wave exhibits a clean $a_2(1320)$-resonance and is included in order to fix the relative phases and stabilize the fit.
In general, there are two components for each wave in the model: a resonance amplitude, \ie a propagator that contains a pole (in this case either the $a_1(1260)$ or the $a_2(1320)$),
and a component with $t$-channel $\pi$ exchange accounting for nonresonant processes.
We parametrize the $a_1(1260)$ propagator by a relativistic Breit-Wigner (BW) amplitude with energy-dependent width saturated by the $\rho\pi$ decay channel~\cite{Akhunzyanov:2018lqa}.
For the resonance part of the $\rho\pi\, D$-wave we employ the $a_2(1320)$ propagator parametrized by a BW amplitude with dynamical width
including the $\rho\pi$ ($80\%$) and $\eta \pi$ ($20\%$) channels, as discussed in~\cite{Akhunzyanov:2018lqa}.
The nonresonant background is added coherently to each wave.
We use an empirical parametrization given by $(m_{3\pi}/m_0-1)^{b}\:\exp[-(c_0+c_1 t'+c_2 t'^2)\tilde{p}^2]$, where $\tilde{p}$ is an effective break-up momentum for the decay into $\xi\pi$ at the given $m_{3\pi}$ value, taking into account the finite width of the isobar $\xi$ and the Bose symmetry of the system, and $m_0=0.5\,\mathrm{GeV}/c^2$ (see Eqs.~(27) and (29) in~\cite{Akhunzyanov:2018lqa}). For the model calculations,
the $t'$ value is fixed to the lower edge of the respective bin.
For the \fnotP-wave, the resonance part of the production amplitude is modified by the $K^\ast\bar{K}\to f_0\pi$-rescattering via the TS.
As the direct decay of the $a_1(1260)$ to the $f_0\pi$ final state has a very slow phase motion and a similar shape as the phenomenological parametrization of the nonresonant part due to the limited phase space (see Supplemental Material for details), the fit cannot distinguish between the two components.
Therefore, this additional component is only considered for systematic studies.

\begin{figure*}[tbp]
    \centering
    \begin{overpic}[width=.325\textwidth]{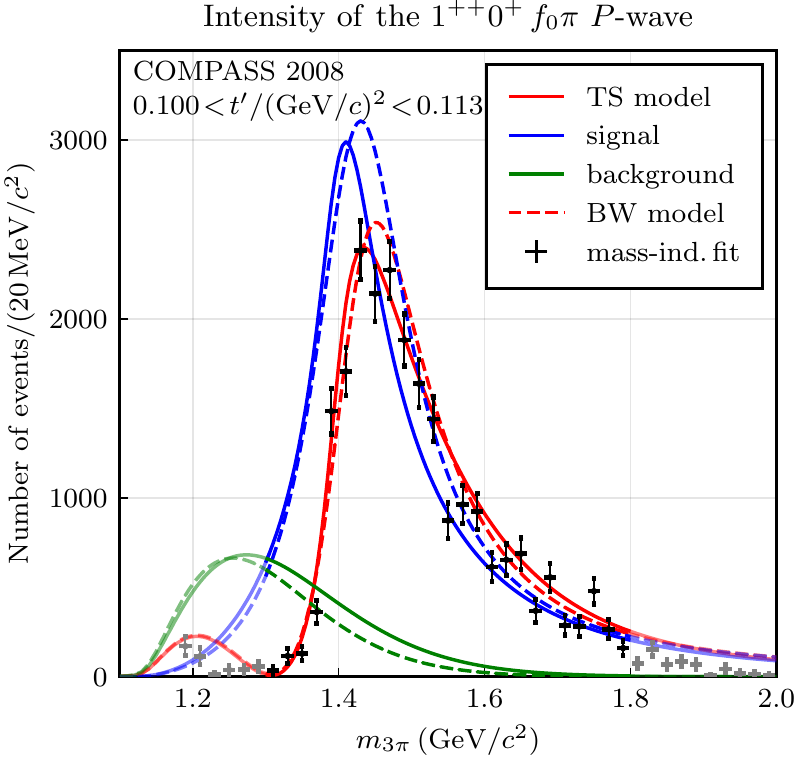} \put (18,75) {(a)}\end{overpic}
    \begin{overpic}[width=.325\textwidth]{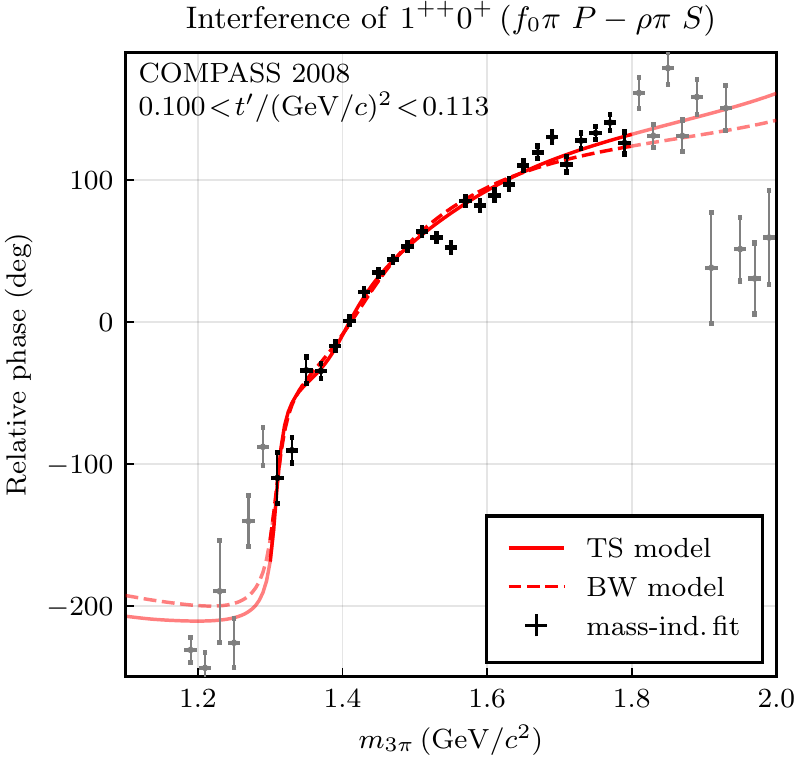} \put (18,75) {(b)}\end{overpic}
    \begin{overpic}[width=.325\textwidth]{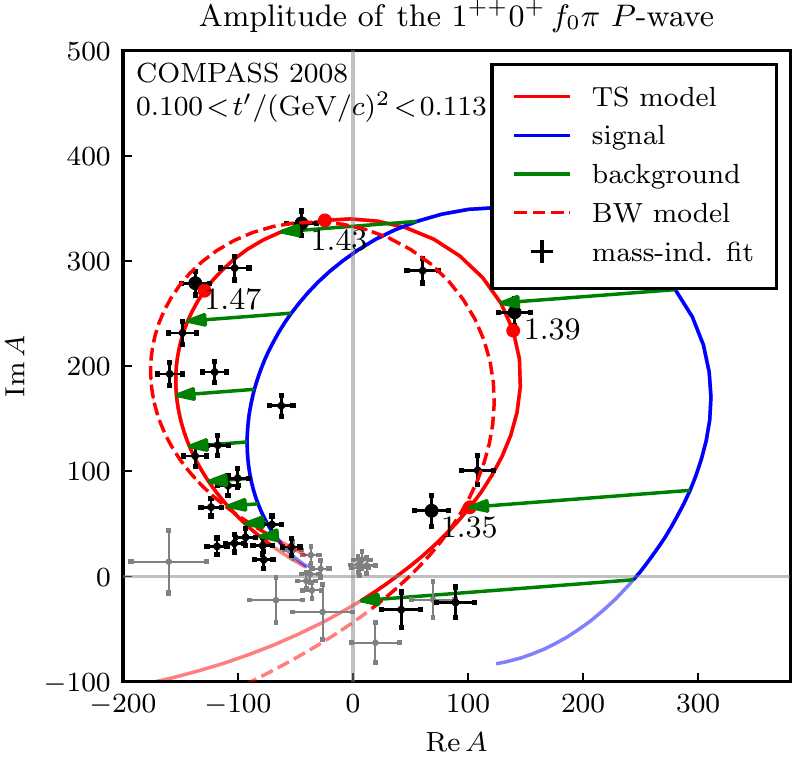} \put (18,75) {(c)}\end{overpic}
    \caption{
    Results of the fit with the TS model (solid lines) and the BW model (dashed lines) to the wave with the $a_1(1420)$ resonance-like signal.
    The fit range is indicated by the color saturation of data points and lines.
    (a) intensity of the \onePPfnotP-wave.
    The complete fit model (red) is decomposed into its signal (blue) and background (green) contributions.
    (b) Relative phase between the  $1^{++}0^{+}\,\rho\pi\,S$-wave and the \onePPfnotP-wave.
    (c) Argand diagram.
    The red dots on the TS-model curve correspond to the indicated $m_{3\pi}$ values in units of $\mathrm{GeV}/c^2$.}
    \label{fig:fit}
\end{figure*}

The free parameters of the model, \ie the $t'$-dependent complex couplings, the background parameters $b$ and $c_i$,
as well as the $t'$-independent BW parameters, are determined by a fit to the COMPASS data points in $m_{3\pi}$ and $t'$ bins.
We note that there are no free parameters influencing the line shape of the TS amplitude, while the strength and the background parameters are adjusted in the fit. As explained in more detail in~\cite{Akhunzyanov:2018lqa},
the data points $y_i$ to be fitted are the intensity and the real and imaginary parts of the interference terms for the 3 selected waves inside the chosen $m_{3\pi}$ ranges (indicated in Fig.~\ref{fig:fit}) for all 11 $t'$ bins.
The fit is performed by
minimizing the sum of the squared differences between data points $y_i$ and model prediction $\hat{y}_i$, weighted by the inverse squared statistical uncertainties:
\begin{align}
    \R^2=\sum_i \frac{(y_i-\hat{y}_i)^2}{\sigma_i^2}.
    \label{eq:Rsq}
\end{align}

Figure~\ref{fig:fit} shows the result of the TS model fit 
in the lowest $t'$ bin, selecting only the \fnotP-wave (full lines).
The fit results for all three waves in all 11 $t'$-bins can be found in~Supplemental Material.
Figure~\ref{fig:fit}(a) shows the intensity of the \fnotP-wave
and Figure~\ref{fig:fit}(b) the relative phase to the \rhoS-wave, both as a function of $m_{3\pi}$.
The resonance-like behavior of the TS amplitude is most evident from the circle in the Argand diagram in panel~(c).
The nonresonant background (green arrows) helps to slightly adjust the position of the circle. Since the phase of the background component does not change with $m_{3\pi}$,
all green arrows are parallel.

In order to evaluate the quality of the TS model fit, we also perform a fit to the data using a simple BW description of the $a_1(1420)$ signal instead of the TS amplitude.
This is accomplished by replacing the 
TS parametrization of the $f_0\pi\, P$-wave 
by a relativistic BW amplitude with free
mass and width parameters assuming the $a_1(1420)$ being a genuine new resonance.
We use a constant-width parametrization since further decay modes of this hypothetical new particle are unknown.
Figure~\ref{fig:fit} shows that the fits with the BW model (dashed) and the TS model (solid) are of very similar quality.
Both models are capable of describing the intensities as well as the corresponding interference terms.
For a quantitative comparison, one can use the quantity defined in Eq.~(\ref{eq:Rsq}).
The biggest contribution comes from the $\rhoS$ and $\rhoD$-waves.
Since the description of these two waves is very similar in both fit models, 
we can omit them for the comparison of the fit quality. In addition, we can exclude one of the two remaining phases of the interferences, since they
depend linearly on one another. Defining $\Rred$ as the reduced weighted sum of the remaining residuals squared divided by the number of degrees of freedom, where only the fit parameters specific to the $\fnotP$-wave are taken into account,
we arrive at a value of $\R^2_{\text{red,TS}} = 4.8$ for the TS and $\R^2_{\text{red,BW}} = 5.2$ for the BW model.
The values indicate that the two fits have comparable quality. The advantage of the TS model is that it has two fit parameters less, since it does not require a new particle with corresponding mass and width.

%%%%%%%%%%%%%%%%%%%%%%%%%%%%%%%%%%%%%%%%%%%%%%%%%%%%%%%%%%
% Systematic Studies
%%%%%%%%%%%%%%%%%%%%%%%%%%%%%%%%%%%%%%%%%%%%%%%%%%%%%%%%%%
To study the stability of the result, we investigate a wide range of sources of systematic uncertainties, both with respect to changes of the model and to changes of the data points.
We perform fits where the data points $y_i$ are varied according to systematic studies for the PWA in bins of mass and $t'$, published in~\cite{Adolph:2015tqa}. 
These include using a smaller wave set, removing negative reflectivity waves, relaxing the event selection,
using a model with relaxed coherence assumption (see~\cite{Adolph:2015tqa} for details) or changing the parametrization of the $f_0(980)$. In an additional study, we use the result of a statistical reanalysis of \cite{Adolph:2015tqa} applying the bootstrap technique~\cite{Efron:1993qfh}.
Also, we consider several variations in the fit model for the TS: (i) a fit with non-Bose-symmetrized phase space;
(ii) neglecting the spins of the particles involved (similar to~\cite{Ketzer:2015tqa});
(iii) including the excitations $a_1(1640)$ and $a_2(1700)$ in the $\rho\pi\,S$ and $\rho\pi\,D$ waves, respectively (mass and width fixed to the values from the PDG~\cite{Tanabashi:2018oca});
and (iv) varying mass and width of the $K^\ast$ resonance according to their uncertainties~\cite{Tanabashi:2018oca} in order to estimate the effect of further rescattering.
The TS model systematically yields a slightly smaller $\R^2_\text{red}$ than the BW model~(see Supplemental Material).

%%%%%%%%%%%%%%%%%%%%%%%%%%%%%%%%%%%%%%%%%%%%%%%%%%%%%%%%%%
% Conclusions
%%%%%%%%%%%%%%%%%%%%%%%%%%%%%%%%%%%%%%%%%%%%%%%%%%%%%%%%%%
In summary, we have shown that the recently discovered resonance-like signal $a_1(1420)$ can be fully explained by the decay of the ground state $a_1(1260)$ into $K^\ast\bar{K}$ and subsequent rescattering through a triangle singularity into the observed final state $f_0(980)\pi$ without the need of a new genuine $a_1$ resonance.
The effect of the triangle singularity, which is expected to be present, is sufficient to explain the observation.

\begin{acknowledgments}
\textit{Acknowledgments.}---We gratefully acknowledge the support of the CERN management and staff as well as the skills and efforts of the technicians of our collaborating institutions.
We would like to thank Bastian Kubis for useful discussions on the scalar form factor.
This work was made possible by the financial support of our funding agencies: MEYS Grant No. LG13031 (Czech Republic); FP7 HadronPhysics3
Grant No. 283286 (European Union); CEA, P2I, and
P.-J. L. was supported by ANR (France) with P2IO
LabEx (ANR-10-LBX-0038) in the framework
"Investissements d'Avenir" (ANR-11-IDEX-003-01);
BMBF Collaborative Research Project 05P2018—
COMPASS, W. D. and M. F. were supported by the
DFG Cluster of Excellence Origin and Structure of the
Universe” (\cite{universe_cluster_de}), M. G. was supported by the DFG
Research Training Group Programmes 1102 and 2044
(Germany); B. Sen Fund (India); Academy of Sciences
and Humanities (Israel); INFN (Italy); MEXT and JSPS,
Grants No. 8002006, No. 20540299, and No. 18540281,
Daiko and Yamada Foundations (Japan); NCN
Grant No. 2017/26/M/ST2/00498 (Poland); FCT Grants
No. CERN/FIS-PAR/0007/2017 and No. CERN/FIS-PAR/
0022/2019 (Portugal); CERN-RFBR Grant No.
12-02-91500, Presidential Grant No. NSh-999.2014.2
(Russia); MST (Taiwan); and NSF (U.S.).
\end{acknowledgments}

%%%%%%%%%%%%%%%%%%%%%%%%%%%%%%%%%%%%%%%%%%%%%%%%%%%%%%%%%%
% References
%%%%%%%%%%%%%%%%%%%%%%%%%%%%%%%%%%%%%%%%%%%%%%%%%%%%%%%%%%
\bibliographystyle{apsrev4-1}
\bibliography{hadron,compass}

\newpage
\onecolumngrid
% \appendix

% \newpage
\section*{Supplemental material} \label{sec:supp}
This supplemental material includes additional information on the systematic studies performed
as well as details of the amplitude calculation. 

\section{Other triangles and the direct decay}
\label{app:others}

\begin{figure*}[ht]
    \centering
    \begin{overpic}[width=.5\textwidth]{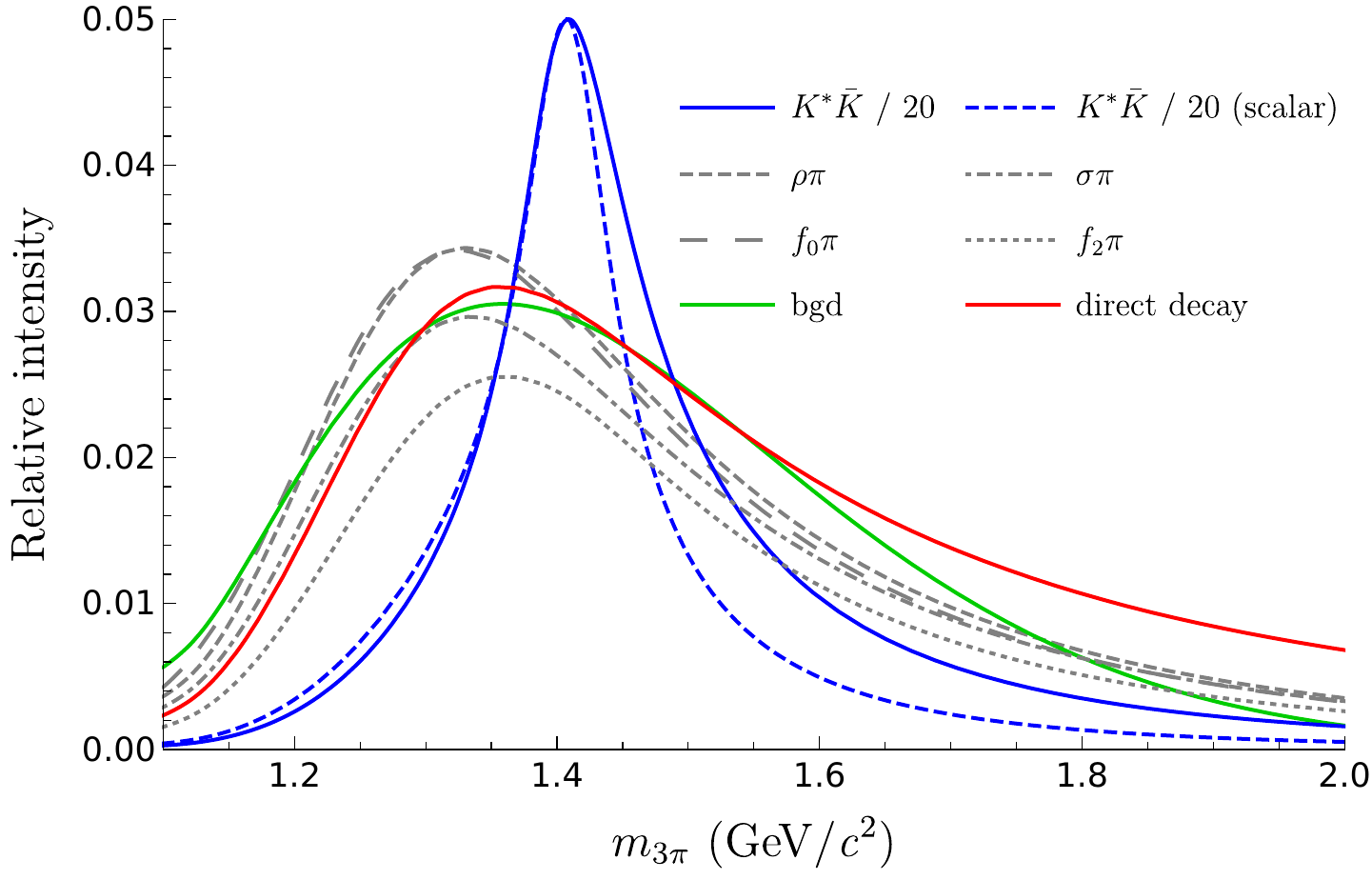} \put (15,60) {(a)}\end{overpic}
    \begin{overpic}[width=.49\textwidth]{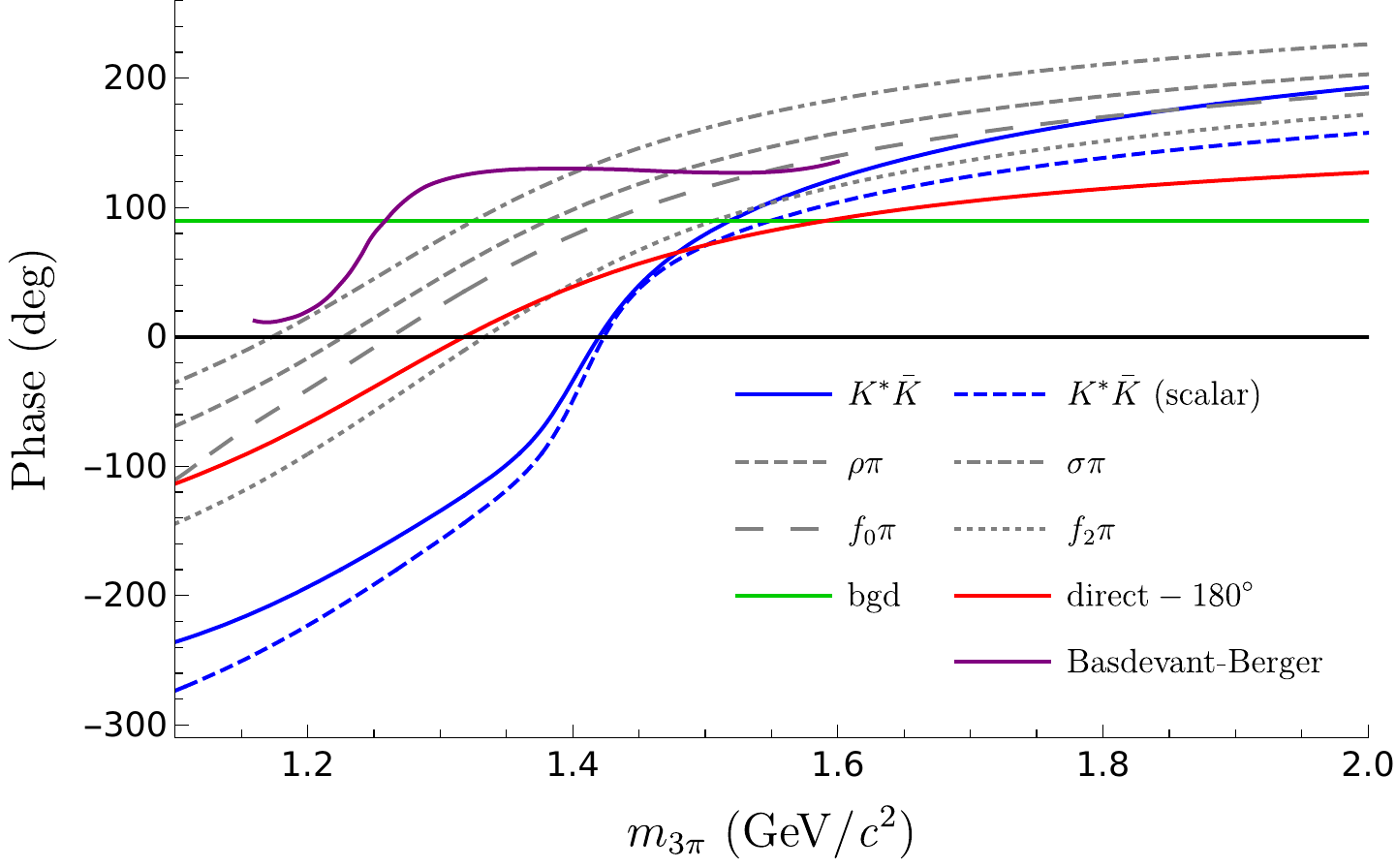} \put (15,60) {(b)}\end{overpic}
    \caption{(a) Intensities and (b) phases of the \fnotP\ amplitude produced from different sources.
    The $m_{3\pi}$ dependence is shown for the $f_0(980)$ subchannel invariant mass fixed to the nominal resonance mass of $990\,\mathrm{MeV}/c^2$.
     See text and Refs.~\cite{Adolph:2015tqa,Akhunzyanov:2018lqa} for details on the calculations and the descriptions of the parameterizations of the isobar and resonance amplitudes. The intensities in %Fig.
     panel~(a) are relative to the maximum intensity of the $K^\ast\bar{K}$ channel, with the couplings in each vertex set to unity. Note that the intensity of the $K^\ast\bar{K}$ triangle graph (blue lines) was scaled down by a factor of $20$. 
    In panel~(b) we also include the phase from Ref.~\cite{Basdevant:2015wma} (purple line). Its inflection point is at a mass of approximately $1.25\,\mathrm{GeV}/c^2$. See text for details on the different curves.
    } 
    \label{fig:triangle}
\end{figure*}

Figure~\ref{fig:triangle} shows the $m_{3\pi}$ dependence of the $f_0(980)\pi$ isobar production amplitude 
for different individual cross channels $\omega$.
The blue lines represent the TS in the $K\bar{K}\pi$ channel with the $K^\ast$ resonance. 
The full blue line is the result of the new partial-wave projection method, Eq.~(1) of the main text, taking into account the spins of all particles involved. 
The dashed blue line labeled ``(scalar)'', as well as all other dashed lines, are obtained by calculations using the Feynman method from Ref.~\cite{Ketzer:2015tqa} assuming that all particles are spinless.
The curves shown in dashed gray correspond to the rescattering effects of the various $\pi\pi$ resonances in the cross channel.
For the calculation we assume that modifications of the lineshapes of the cross-channel resonances due to rescattering are negligible.
It can be seen that the $K^\ast\bar{K}$ channel produces a narrow peak and a strong phase motion at the mass of the $a_1(1420)$ due to the TS being very close to the physical region, while all other channels including the direct decay and the non-resonant background (Bgd) manifest themselves in a broad bump and a slow phase motion.
As the direct decay of the $a_1(1260)$ to the $f_0\pi$ final state has a very slow phase motion and a similar shape as 
the phenomenological parameterization of the non-resonant part (compare the red and green curves in Fig.~\ref{fig:triangle}), the fit cannot distinguish between the two components. The direct decay of the $a_1(1260)$ to $f_0\pi$ can hence be omitted.

\section{Systematic studies}
\label{app:syst}
\begin{figure*}[ht]
\centering
\includegraphics[width=.65\linewidth]{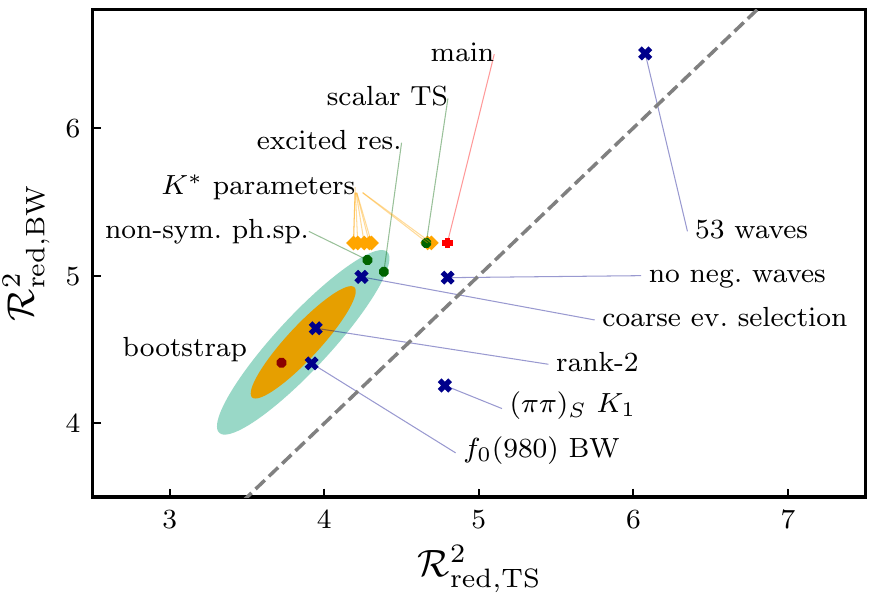}
\caption{
    $\mathcal{R}_{\mathrm{red,BW}}^2$ for the Breit-Wigner model vs.\ $\mathcal{R}_{\mathrm{red,TS}}^2$ for the TS model.
    See main text for details on their definition. The main fit shown in Fig.~2 in the main text, is represented by the red cross, the gray dashed line indicates $\mathcal{R}_{\mathrm{red,BW}}^2 = \mathcal{R}_{\mathrm{red,TS}}^2$.
    Blue crosses correspond to systematic studies using different data points and green dots
    show the fit results with a modified model of the $\fnotP$ wave.
    Fits with a modified lineshape of the $K^*$ resonance are shown by the orange diamonds.
    The result of the bootstrap analysis is shown by the filled ellipses
    which cover $68\,\%$, and, $95\,\%$ of the obtained sample, respectively;
    the fit to the bootstrap-sample mean is represented by the brown point.}
\label{fig:systematics.chi2comparison}
\end{figure*}

We investigate several sources of systematic uncertainties including variations of the fit model and uncertainties of the mass-independent PWA.
We perform fits where the COMPAS data points of the main text are varied according to systematic studies for the PWA in bins of mass and $t'$ (see Ref.~\cite{Adolph:2015tqa}). 
Figure~\ref{fig:systematics.chi2comparison} compares the quantity $\Rred$, as defined in Eq.~(2) of the main text, for the TS model and the model using a simple Breit-Wigner amplitude for the $a_1(1420)$, for the various studies. 
These include using a smaller wave set (``53 waves''), removing negative-reflectivity waves (``no neg.\ waves''), relaxing the event selection (``coarse ev.\ sel.''), using rank 2 instead of rank 1 (``rank 2''), and changing the parameterization of the $f_0(980)$ (``$f_0(980)\,\mathrm{BW}$'', ``$(\pi\pi)_S\ K_1$''). In an additional study, we use the result of a statistical re-analysis of \cite{Adolph:2015tqa}  applying the bootstrap technique~\cite{Efron:1993qfh} (``bootstrap'').
Also, we consider several variations of the fit model for the TS: (i) a fit with non-Bose-symmetrized phase space (``non-sym.\ ph.\ sp.''); (ii) neglecting the spins of the particles involved (similar to Ref.~\cite{Ketzer:2015tqa}, ``scalar TS''); (iii) including the excitations $a_1(1640)$ and $a_2(1700)$ in the $\rho\pi\,S$ and $\rho\pi\,D$ waves, respectively (masses and widths fixed to the values from the PDG~\cite{Tanabashi:2018oca}, ``excited res.'');
and (iv) we estimate the effect of further rescattering by varying mass and width of the $K^\ast$ resonance according to their
uncertainties~\cite{Tanabashi:2018oca} (``$K^*$ parameters'').
We see from Fig.~\ref{fig:systematics.chi2comparison} that the TS model systematically yields a slightly smaller $\R^2_\text{red}$ than the Breit-Wigner model.

\section{The cross-channel projection integral}\label{sec:calculation}
The contribution to the dispersion term for the $J^{PC}=1^{++}\,f_0\,\pi\,P$-wave in Eq.~(1) that manifests the peaking structure from TS is the $K^*\bar{K}\,S$-wave. It reads:
\begin{align}
  F_{f_0\pi\,P}^{\{1\}}(s,\sigma_{\{1\}}) = k(s,\sigma_{{\{1\}}}) \bigg( C_W^{\{1\}}(s) 
    + \frac{1}{2\pi} \int_{4m_K^2}^\infty \frac{\rho_{K\bar{K}}(\sigma) \hat F_{f_0\pi\,P,K^*\bar{K}\,S}^{\{1\}}(s,\sigma)}{k(s,\sigma)(\sigma - \sigma_{\{1\}}-i\varepsilon)}\,\diff\sigma \bigg) \frac{1}{\sqrt{1+k^2(s,\sigma_{{\{1\}}}) R^2}}\,,
\end{align}
where $k(s,\sigma) = \sqrt{\lambda(s,\sigma,m_\pi^2)/(4s)}$,
the phase space factor $\rho_{K\bar{K}}(\sigma) = \sqrt{\lambda(\sigma,m_K^2,m_K^2)}/(8\pi \sigma)$,
and $\lambda$ is the K\"all\'en function, $\lambda(x,y,z) = x^2+y^2+z^2-2(xy+yz+zx)$.
To limit the growth of the $P$-wave amplitude, we add a customary Blatt-Weisskopf barrier factor~\cite{Blatt:1952,VonHippel:1972fg} with the size parameter $R=5\,$GeV$^{-1}$.
The projection integral reads:
\begin{align}
  \hat F_{f_0\pi\,P,K^*\bar{K}\,S}^{\{1\}}(s,\sigma_1) = \frac{\sigma_1}{\lambda(s,\sigma_{1},m_\pi^2)\sqrt{\lambda(\sigma_1,m_K^2,m_K^2)}}
  \int_{\sigma_3^-(s,\sigma_1)}^{{\sigma_3^+(s,\sigma_1)}}
  \frac{Q(\sqrt{s},\sqrt{\sigma_1},\sqrt{\sigma_3})}
    {(\sqrt{\sigma_3}+\sqrt{s}-m_3)(\sqrt{\sigma_3}+\sqrt{s}+m_3)}
    \frac{\diff\sigma_3}{D_{K^\ast}(\sigma_3)}\,,
\end{align}
where we integrate over the phase space of the $K\bar{K}\pi$ system with $\sigma_3 = m^2(K\pi)$ and $\sigma_1 = m^2(K\bar{K})$. The integration limits as calculated as follows:
\begin{align}
    \sigma_3^\pm(s,\sigma_1) = m_K^2 + m_\pi^2 + (s-\sigma_1-m_\pi^2)/2 \pm \sqrt{\lambda(s,\sigma_1,m_\pi^2)\lambda(\sigma_1,m_K^2,m_K^2)}/(2\sigma_1)\,.
\end{align}
Beyond the physical region, we use analytic continuation with the $s\to s+i\epsilon$ prescription~\cite{Bronzan:1963mby}.
The function $Q(x,y,z)$ is polynomial in $x$, $y$, and $z$ and arises from the product of Wigner d-functions (see Eq.~(B12) in Ref.~\cite{Mikhasenko:2018bzm}),
\begin{align}
    Q(x,y,z) &= 
    m_{1}^{4} m_{3}^{2} + m_{1}^{4} x^{2} + 2 m_{1}^{4} x z + m_{1}^{4} z^{2} - m_{1}^{2} m_{2}^{2} m_{3}^{2} - 3 m_{1}^{2} m_{2}^{2} x^{2} - 2 m_{1}^{2} m_{2}^{2} x z\\ \nonumber
    &\quad + m_{1}^{2} m_{2}^{2} z^{2} + m_{1}^{2} m_{3}^{2} x^{2} + 4 m_{1}^{2} m_{3}^{2} x z - m_{1}^{2} m_{3}^{2} y^{2} + m_{1}^{2} m_{3}^{2} z^{2} - 2 m_{1}^{2} x^{3} z\\ \nonumber
    &\quad + m_{1}^{2} x^{2} y^{2} - 2 m_{1}^{2} x^{2} z^{2} - 2 m_{1}^{2} x y^{2} z - 2 m_{1}^{2} x z^{3} - m_{1}^{2} x^{4}\\ \nonumber
    &\quad - 3 m_{1}^{2} y^{2} z^{2} - m_{1}^{2} z^{4} - m_{2}^{2} m_{3}^{2} x^{2} + m_{2}^{2} m_{3}^{2} y^{2} + m_{2}^{2} x^{4} - m_{2}^{2} x^{2} y^{2}\\ \nonumber
    &\quad - 3 m_{2}^{2} x^{2} z^{2}+ 2 m_{2}^{4} x^{2} + 2 m_{2}^{2} x y^{2} z - 2 m_{2}^{2} x^{3} z - m_{2}^{2} y^{2} z^{2} + m_{3}^{2} x^{2} z^{2}\\ \nonumber
    &\quad - m_{3}^{2} y^{2} z^{2} + x^{4} z^{2} - 3 x^{2} y^{2} z^{2} + x^{2} z^{4} - 2 x y^{2} z^{3} + 2 x^{3} z^{3} + 2 y^{4} z^{2} + y^{2} z^{4}\,,
\end{align}
with $m_1=m_\pi$, and $m_2=m_3=m_K$.
The term $D_{K^*}(\sigma_3)$ is the denominator of the $K\pi$ scattering amplitude, \ie
\begin{align}
    D_{K^*}(\sigma_3) = m_{K^*}^2-\sigma_3-m_{K^*} \Gamma_{K^*} (\Sigma_{K^\ast}(\sigma_3) - \text{Re}\,\Sigma(m_{K^\ast}^2)) / \text{Im}\,\Sigma_{K^\ast}(m_{K^\ast}^2)\,.
\end{align}
We ensure the correct analytic structure using the Chew-Mandelstam function $\Sigma(\sigma_3)$ for the absorptive term~\cite{Tanabashi:2018oca,Basdevant:1977ya}.
Note that $\text{Im}\,\Sigma(\sigma_3) = \sqrt{\lambda(\sigma_3,m_\pi^2,m_K^2)}/(16\pi\sigma_3)$.

\section{Fit result for all $t'$ bins}
\begin{figure}[tbph!]
    \centering
    \includegraphics[width=\textwidth,page=1]{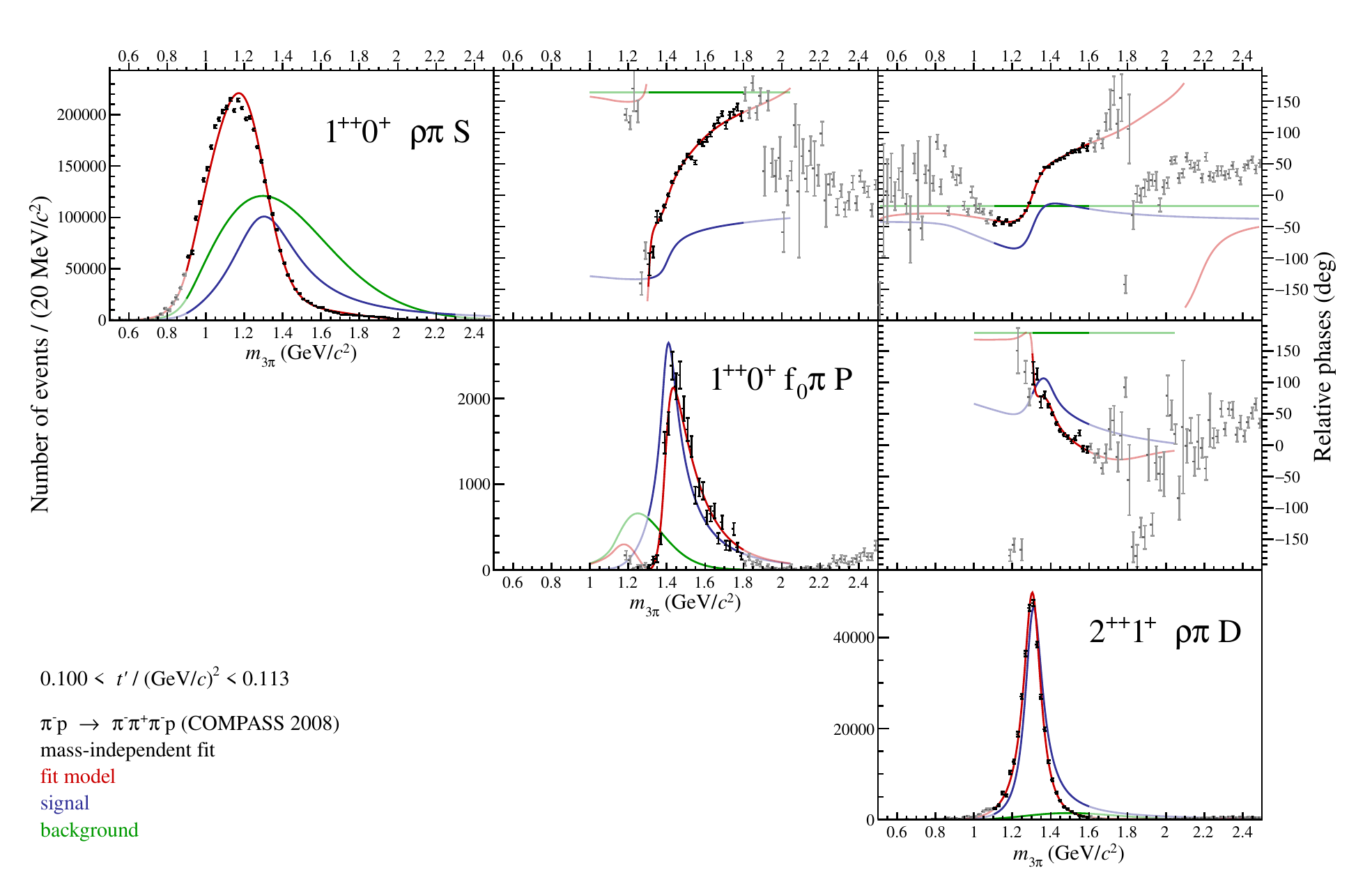}
    \caption{Spin-density matrix elements of the 3 waves selected for the TS model fit and the corresponding fit results, both shown for the first (lowest) bin of $t^\prime$. 
    The SDMEs as a function of $m_{3\pi}$ are visualized in the form of a $3 \times 3$ upper-triangular matrix of graphs with the partial-wave intensities on the diagonal and the relative phases between the partial waves on the off-diagonal.
Black crosses correspond to the result of the PWA in bins of $m_{3\pi}$ and $t'$ from Ref.~\cite{Adolph:2015tqa} with statistical uncertainties indicated by vertical lines. The data are overlaid by the TS model curve (red), the contributions from signal (blue) and non-resonant background (green). The $m_{3\pi}$ fit range is indicated by the color saturation of data points and fit result. Regions
indicated by lower color saturation were not included in the fit; the model curves in these regions are extrapolations.}
    \label{fig:stampplot1}
\end{figure}

Figures~\ref{fig:stampplot1} -- \ref{fig:stampplot11} show the spin-density matrix elements (SDMEs) of the three waves selected for the model fit (data points) and the fit results of the TS model for 
all $t'$-bins. 
The full fit model (red) is decomposed into signal (blue) and background (green) as described in the main text. The intensities are plotted on the diagonal and the complex phase of the interference parts on the off-diagonal. The 3 rows as well as the 3 columns correspond to the $1^{++}0^+ \rho\pi S$, $1^{++}0^+ f_0\pi P$, and $2^{++}1^+ \rho\pi D$-waves.

{}

\foreach \n in {2,...,11}{
\begin{figure}[tbph!]
    \centering
    \includegraphics[width=\textwidth,page=\n]{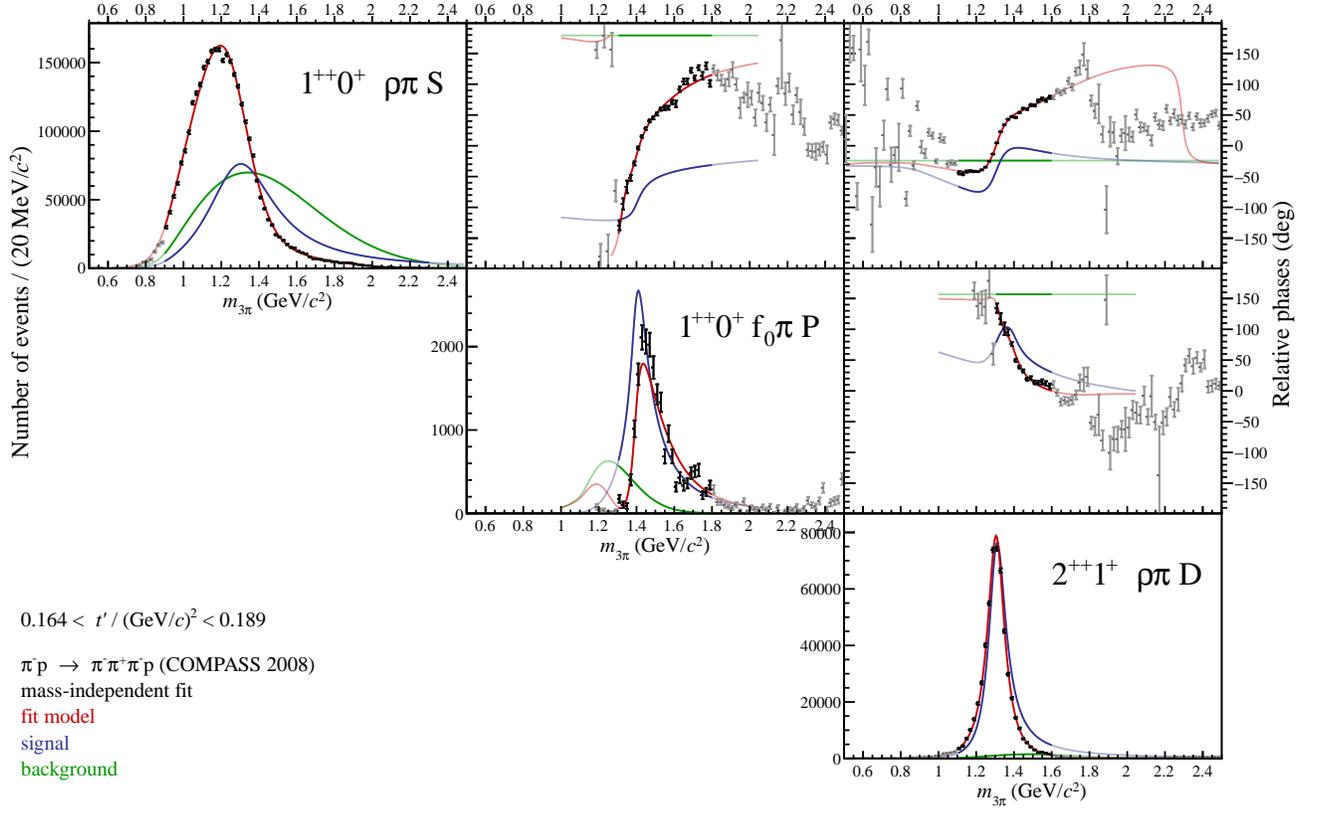}
    \caption{Same as Fig.~\ref{fig:stampplot1}, but for $t^\prime$-bin \n.}
    \label{fig:stampplot\n}
\end{figure}{}
}

\end{document}